\documentstyle[aps,epsf,epsfig,prl,psfig]{revtex}

\newcommand{\avg}[1]{\langle{#1}\rangle}

\newcommand{\be}{\begin{equation}\FL}
\newcommand{\ee}{\end{equation}}
\newcommand{\beas}{\begin{eqnarray*}}
\newcommand{\eeas}{\end{eqnarray*}}
\newcommand{\bea}{\begin{eqnarray}\FL}
\newcommand{\eea}{\end{eqnarray}}
\newcommand{\ovl}[1]{\overline{#1}}

\def\erf{\hbox{erf}\,}

\begin{document}

\twocolumn[\hsize\textwidth\columnwidth\hsize\csname
@twocolumnfalse\endcsname

\title{From Minority Games to real markets}
\author{D. Challet${}^{(1)}$, A. Chessa${}^{(2)}$,
M. Marsili${}^{(3)}$ and Y.-C. Zhang${}^{(1)}$} 
\address{${}^{(1)}$
Institut de Physique Th\'eorique, Universit\'e de Fribourg, 
Perolles CH-1700\\
${}^{(2)}$ Dipartimento di Fisica and Unit\'{a} {\it INFM}, Universit\'{a} di
Cagliari, I-09124 Cagliari\\
${}^{(3)}$ Istituto Nazionale per la Fisica della Materia ({\it INFM}),
Unit\'a di Trieste-SISSA, I-34014 Trieste}
\date{\today} 
\maketitle 
\widetext
 
\begin{abstract}
We address the question of market efficiency using the Minority Game
(MG) model. First we show that removing unrealistic features of the MG
leads to models which reproduce a scaling behavior close to what is
observed in real markets. In particular we find that {\em i)} fat
tails and clustered volatility arise at the phase transition point and
that {\em ii)} the crossover to random walk behavior of prices is a
finite size effect. This, on one hand, suggests that markets operate
close to criticality, where the market is marginally efficient. On the
other it allows one to measure the distance from criticality of real
market, using cross-over times. 
The artificial market described by the MG is then studied as an
ecosystem with different {\em species} of traders. This clarifies 
the nature of the interaction and the particular role played by
the various populations.
\end{abstract}

\pacs{PACS numbers: 02.50.Le, 05.40.+j, 64.60.Ak, 89.90.+n}
\narrowtext

]

\section{Introduction}

The Minority Game\cite{CZ1,web} (MG) was initially designed as the
most drastic simplification of Arthur's famous El Farol's Bar problem
\cite{Arthur}: it describes a system where many heterogeneous agents
interact through a price system they all contribute to determine. The
MG is an highly stylized model of such a situation: it captures some
key features of a generic market mechanism and the basic interaction
between agents and public information -- i.e. how agents react to
information and how these reactions modify the information itself. In
addition, it allows to study in details how macroscopic quantities
depend on microscopic behaviors.

However, the basic MG is a so stylized model of a financial market
that prices are not even explicitly defined. Furthermore the
micro-economic behavior of agents is quite simplified: agents have
heterogeneous strategies but they enter the game with the same
weight. In other words there are not poorer or richer agents and their
wealth does not change according to their performance. Also all agents
are constrained to play, with the same frequency, no matter how much
they may loose. All these unrealistic features makes it hard to accept
the MG as a model of a real financial market, especially when compared
to other agent-based models
\cite{LuxMarchesi,Caldarellietal,Solomon,ContBouchaud} which have so
far been more successful in reproducing the stylized facts of high
frequency statistics of prices \cite{MantegnaStanley}.

The same stylized nature of the MG however, allows one to gain a deep
understanding of its extremely rich collective behavior: Statistical
mechanics of disordered systems indeed allows for a full analytic
solution in the limit of infinitely many agents \cite{MCZ}.  More
precisely, these techniques allows one to fully characterize the
evolutionary equilibrium of the dynamical learning process in a truly
complex system of interacting adaptive agents. In a top-down approach
to real financial markets, where complexity is added in steps, the
analytic solution of the MG provides an invaluable starting point
which allows us to keep full control on the emergent features. Several
extensions in this direction where discussed elsewhere
\cite{MMM,DeMM,cavagna2}.

The purpose of the present paper is to advance even more in this
endeavor. First we show that, by removing further unrealistic
features, and defining a price process in terms of excess demand, the
main stylized facts of high frequency price fluctuations are recovered
within the MG. In particular we allow agents to have different weights
in the market according to their wealth, which evolves as a result of
their trades\footnote{Other authors also considered this extension 
\cite{FarmerJoshi,JohnLiege}. See section \ref{cap}}. As in the MG we find a phase transition between a
symmetric (information-efficient) phase and an asymmetric phase,
depending on the ratio $\alpha$ between agents and information
complexity.  The symmetric phase, in this case, is characterized by
zero excess demand and constant prices and is hence similar to an
absorbing phase. Statistical features such as fat tails in the
distribution of returns and long-time volatility autocorrelation, only
arise close to the critical point $\alpha_c$.

Second, we derive a coherent picture of the collective behavior of a
market. In this picture, we can regard a market as an ecology of
different ``species'' of investors, each playing his particular
role. On one side there are traders who need the market for exchanging
goods and are not interested in speculation. These kind of agents --
called {\em hedgers} in the economic literature \cite{Working} -- will
be called {\em producers} hereafter, following ref. \cite{MMM}. On the
other, there are bounded rational agents {\em speculators} equipped
with inductive thinking and very heterogeneous strategies, acting as
scavengers of information.  We can offer a coherent picture of how the
resulting food chain operates: Producers inject a limited amount of
information, upon which a swarm of speculators feed. The two groups in
the market ecology have only partial overlap in interest.  This calls
for two parameters to characterize efficiency because market
efficiency is interpreted differently from different
players. Producers would like that the information content is small,
and fluctuations also. Whereas speculators would like small
fluctuations, but they prefer when the information content is large.
%
%We further study how the quality of the
%speculators influences the market efficiency. We find that better
%performing speculators helps to reduce fluctuations, thus play a
%positive role toward the community; pure noise traders do harm to
%themselves as well as to others; while the worst performing players
%can make damages worst than the noise traders (?????????).

Thus the MG provides a coherent picture of how markets function which,
on one hand is rooted on an analytic approach providing deep insights
on the collective statistical behavior and, on the other, is able to
reproduce the main statistical regularities -- the so-called stylized
facts -- of financial markets.

We keep our discussion as simple and informal as possible.  Formal
mathematical definitions and technical details can be found in the
appendix.

\section{The MG as a coarse grained model of a market}

Naively speaking, what agents do in a financial market is to gather
information on the present state of the market and to process it in
order to determine an investment strategy. We call this mapping from
information to action a {\em trading strategy}. One can regard a
market as an ``evolutionary soup'' of trading strategies competing
against each other\cite{ZENews,FarmerEco}. Modeling this system is a
quite complex task: First because trading strategies, in general, live
in a very complex and high dimensional functional space (specially
because of their inter-temporal nature).
Secondly, trading strategies involve all sorts of details --
such as expectations, beliefs, how agents behave under uncertainty and
how they discount the future -- which are heterogeneous across agents.
Finally time constraints and information or computational complexity
may induce agents to a sub-optimal, boundedly rational behavior
\cite{Arthur}.

This situation forces one either to models whose complexity is of the
same order of reality, and that are hence useless, or to work under
some simplifying assumptions. The MG is based on
the following simplifications\footnote{For a more 
detailed definition of the MG we refer the reader to 
ref. \cite{CM,CMZe,MMM,MCZ} as well as to the appendix.}:
\begin{enumerate}
\item time is discrete, i.e. market interaction is repeated for 
a infinitely many periods.
\item information is discretized in one of $P$ events labeled by an
integer $\mu$, which is drawn at random, independently in each 
period\cite{cavagna}.
\item actions are discretized in a binary choice $a_i(t)\in \{-1,1\}$
at each period $t$ for each agent $i$
\item The space of trading strategies is then the set of
binary functions $f:(1,\ldots,\mu,\ldots,P)\to (-1,+1)$.
\item Agents are heterogeneous: Each agent is endowed 
with a finite number $S$ of trading strategies, 
which are drawn at random and 
independently for each agent from the set of all possible strategies.
\item Agents are adaptive: they evaluate the performance of their
strategies while using their best one for actual trading.
The adaptive process is similar reinforcement learning dynamics 
\cite{rust,CamererHo} but traders behave non-strategically, i.e.
as price-takers (see ref. \cite{MCZ} for more details).
\item the market mechanism is a minority game: Agents who took the
minority action are rewarded whereas the majority of agents loses. This 
captures the fact that markets are mechanisms for reallocation of 
resources so that no gain is possible, in principle, by pure trading.
If some agent gains, some other must lose.
With $A(t)=\sum_i a_i(t)$ being the sum of individual actions $a_i(t)$,
a simple choice of payoffs of minority type is
\be 
u_i(t)=-a_i(t)A(t).  
\ee
If $A(t)>0$, traders who took $a_i(t)=-1$ win, whereas those who
took $a_i(t)=+1$, which are the majority, loose.
\end{enumerate}

This is a coarse grained description of a market in the sense that it
does not enter into the details of the behavior of agents nor of the
market mechanism. Both are considered as {\em black boxes} containing
all sorts of complications. We just retain the key features of
{\em i)} heterogeneity and bounded rationality for agents, and 
{\em ii)} the minority nature for the market mechanism. 

It needs to be said that such a coarse description also requires an
abstraction of usual terms such as prices, volume and excess demand at
a more generic level. For example it is natural to relate $A(t)$ --
which is the unbalance between the two group of agents -- to the
excess demand. Indeed the latter measures in a real market the
unbalance between buyers and sellers. 
In view of the statistical nature of the laws which govern the
collective behavior, and of the robustness of these laws with respect
to changes in microscopic details, such a stretch of the customary meaning 
of common economic terms may be justified.

It is also worth stressing that there
is no {\em a priori} best trading strategy in the market depicted by
the Minority Game. This justifies the equiprobability assumption by
which strategies are drawn. Whether a strategy is good or bad cannot
be decided {\em a priori}; rather the quality of a strategy depends on
how it will perform against the other strategies present in the
market.

The two features discussed above are enough to reproduce a remarkably
rich behavior: The key variable is the ratio $\alpha=P/N$ between
information diversity $P$ and the number $N$ of agents
\cite{savit}. The collective behavior is characterized by market's
predictability $H$\footnote{$H$ is not the only measure of predictability, but is the only one relevant for standard agents. Different agents can profit from other types of predictability \cite{MMM}.},  and global efficiency $\sigma^2$ (see appendix I). The first ($H$)
measures how the market outcome $A(t)$ is correlated with the
information $\mu(t)$, i.e. whether a positive $A(t)$ is more or less
likely than a negative one when the information is $\mu$. $H>0$
implies that knowledge of $\mu$ allows some prediction of the sign of
$A(t)$; accordingly some agents have a positive gain.  The second 
($\sigma^2=-\sum_i \avg{u_i}$) is related to the
total loss suffered by agents, which means that the MG is a negative sum game.
 When few agents are present (large
$\alpha$) the market is easily predictable (i.e. $H>0$) and agents
perform only slightly better than random agents (who decide their
actions on the basis of coin tossing). As more and more agents are
added, the market becomes more efficient both because agents payoffs
increase on average (i.e. $\sigma^2$ decreases) and because the market
becomes less predictable (i.e. $H$ decreases).  A phase transition
takes place \cite{savit,CM,CMZe} at a critical value $\alpha_c$, where
agents average gain reaches a maximum and the market's outcome becomes
unpredictable \cite{CM} ($H=0$). Below this value of $\alpha$ the
market remains unpredictable ($H=0$) and the losses of agents
($\sigma^2$) increase in a way which is specially dramatic if agents
are very reactive \cite{MC}. All these features generalize to a number
of situations\footnote{A qualitatively different behavior arises,
instead, if agents abandon the price taking behavior and account for
their market impact. Assuming that agents behave as price takers, we
shall not discuss this case here but rather refer the interested
reader to refs. \cite{MCZ,DeMM,CMZe,MC}.} such as {\em i)} including a
fraction of deterministic agents -- so called {\em producers} or {\em
hedgers} \cite{Working} -- who have but one strategy and can make the market a positive sum game\cite{MMM}, {\em
ii)} allowing for some correlation in the pool of strategies held by
each agent \cite{MMM}, or {\em iii)} allowing for agents with
different constant weights. The phase diagram for this last case is
reported in Fig. \ref{figw} (see the appendix for details on the
calculation).

\begin{figure}
\centerline{\psfig{file=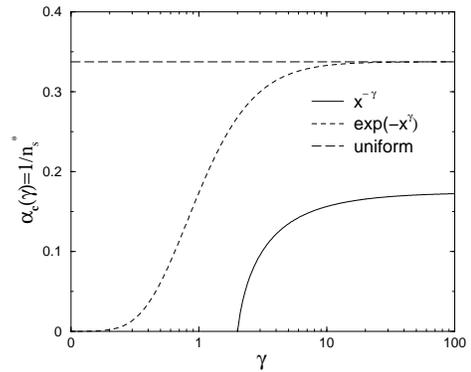,width=6cm}}
\caption{Phase diagrams of the MG with weighted agents for different 
distributions of weights $w_i$: 
power law (full line) and stretched exponential (dotted line) 
distribution with parameter $\gamma$. The results for a 
uniform distribution is also shown as a dashed line. Below the line
corresponding to each distribution, 
the market is in the symmetric phase $H=0$.}
\label{figw}
\end{figure}

The emergent picture is that when agents are few, the market is rich of 
profitable trade opportunities. These may attract other agents in the
market. As the number of agents increases, these opportunities are 
eliminated and the market is driven towards information efficiency 
($H=0$). This suggests that the real markets should operate close 
to the critical point $\alpha_c$ where profitable trade opportunities
are barely detectable. The process by which the market self-organizes 
close to the critical point is more likely to be of evolutionary nature
and hence to take place on longer time-scales\footnote{This indeed agrees 
with the fact that self-organized criticality generically arises in system 
where the dynamics leading to internal re-organization of the system 
occurs on a much faster time-scale than that characterizing the dynamics
of external perturbations (see e.g. \cite{SOC}).}

There are two unnatural features in the MG at this stage: 
First agents are always constrained to play, even if they lose a lot, 
and second the performance of an agent does not affect his wealth. In
reality each trader is allowed only to lose a finite amount of money,
after which it goes bankrupt and exits the game. In the following we 
shall see that the correction to these two shortcomings leads indeed
to quite realistic results \footnote{See also \cite{JohnLiege}.}.

\section{The MG with dynamical capitals}
\label{cap}
How is this scenario modified if one accounts for the fact that agents
have a fixed budget $c_i$ which itself evolves as a result of their
trading? We address this issue by making $c_i(t)$ -- the capital held
by agent $i$ at time $t$ -- a dynamical variable and assuming that
each agent $i$ invests a fraction $\epsilon$ of it in the market.
Speculators have no other gain than that resulting from trading, so
that $c_i(t)$ evolves as a result of it. On the other hand, producers
-- who have other revenues and use the market for reallocation of
resources -- always invest a fixed quantity (see the appendix for more
details). In a loose sense the model becomes evolutionary. 
Indeed poorly performing strategies lead to capital losses and are 
therefore washed out of the market. On the other hand good strategies 
imply capital increase, which enhances the negative effects of market
impact. As a result capitals adjust in order to balance strategy's 
performance and market impacts. 

Similar models with dynamical capitals, based on the minority game,
have been studied in refs. \cite{FarmerJoshi,JohnLiege}. Ref.
\cite{FarmerJoshi} expands in much details on the micro-economics of
these types of models and explores how collective behavior depends on
it. The agent based models discussed in ref. \cite{JohnLiege} pay also
considerable attention to realism at the micro-economic level.  The
price for this is that one needs to introduce many parameters which
implies that one looses the contact with picture provided by the
analytic solution to the MG\cite{CMZe}. Our approach is instead based
on this picture and aims primarily at establishing what elements of
this picture persist when the complexity of the model increases.  Key
questions, for us, are whether the phase transition is robust to such
changes and whether anomalous scaling of price returns
\cite{MantegnaStanley} are related to the critical point or not.
As we shall see, we find positive answers in both cases, which 
open a new perspective on market's efficiency.

\begin{figure}
\centerline{\psfig{file=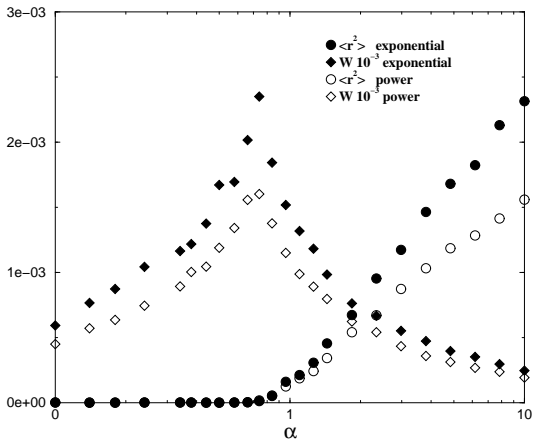,width=6cm,angle=270}}
\caption{Behavior of the MG with dynamic capitals as a function of
$\alpha$: fluctuation of returns $\avg{r^2}$ ($\circ$) and average
wealth of speculators ($\diamond$). The system is composed of $50$
producers and $50$ speculators.  Simulations were carried out for a
number of time steps large enough to reach a stationary state (which 
typically requires $10^6$ time steps). The
results depend very weakly on $\epsilon$, as long as it is small, and on
the distribution of wealth of producers:
A power law distribution with exponent $3/2$ (open symbols) yields
very similar results to those obtained with an exponential
distribution (full symbols).}
\label{figw1}
\end{figure}

The results of numerical simulations, as a function of $\alpha$ are
shown in figure \ref{figw1}. As the information complexity $P$
decreases (or as number of agents increases) i.e. as $\alpha$
decreases, the market becomes less and less predictable. Again at a
critical value $\alpha_c$ the market becomes unpredictable. Actually
the dynamics of $c_i(t)$ reaches a point where the return $r^\mu$ to
the investment under information $\mu$ vanishes for all
$\mu=1,\ldots,P$. Hence the phase $\alpha<\alpha_c$ is an {\em
absorbing} phase where no dynamics actually takes place. The
statistical properties of the stationary state are in principle
accessible to an analytic approach along the lines of
refs. \cite{CMZe,MCZ} for $\epsilon\ll 1$ as discussed in the
appendix.  Interestingly, $\alpha_c$ marks also the point where the
relative wealth of speculators is maximal, as can be seen in fig.
\ref{figw1}. The distribution of wealth across agents falls off
exponentially, with a characteristic wealth which is maximal at
$\alpha_c$.

\begin{figure}
\centerline{\psfig{file=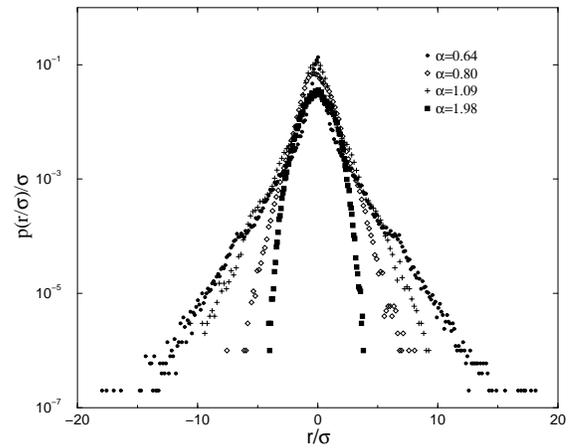,width=6cm,angle=270}}
\caption{Probability density function (pdf) of returns (rescaled to
unit variance) for $\alpha=0.64=\frac{260}{401}$,
$0.80=\frac{240}{301}$, $1.09=\frac{220}{201}$, 
$1.98=\frac{200}{101}$. Tails become fatter and fatter as the critical
value $\alpha_c\approx 0.6$ is approached.}
\label{figpdfw}
\end{figure}

\begin{figure}
\centerline{\psfig{file=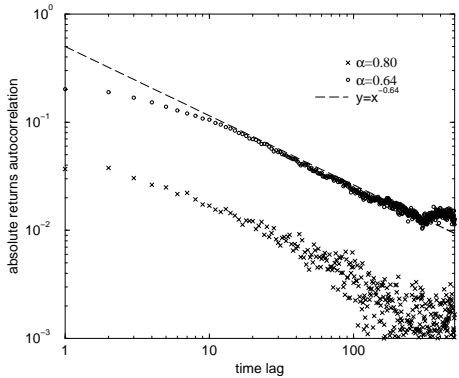,width=6cm}}
\caption{Volatility correlation function
$\avg{|r(t)||r(0)|}-\avg{|r(0)|}^2$ of absolute returns for
$\alpha=0.64=\frac{260}{401}$, $0.80=\frac{240}{301}$,
. Exponent of the autocorrelation function is about -0.64
for $\alpha=0.64$}
\label{figvolcls}
\end{figure}

Having defined returns from trading, allows one to define a price
$p(t)$ as the sum over time of returns (see the appendix for more
details).  Then one can investigate the statistical properties of the
price time series and compare it to empirical findings
\cite{MantegnaStanley}.  Remarkably figure \ref{figpdfw} shows that
fat tails similar to those observed in real markets emerge close to
the critical point $\alpha_c$. In addition figure \ref{figvolcls}
shows that volatility clustering also emerges close to the critical
point: The correlation function of absolute values of returns has an
algebraic decay with time close to $\alpha_c$ which turns into an
exponential decay away from criticality.

\begin{figure}
\centerline{\psfig{file=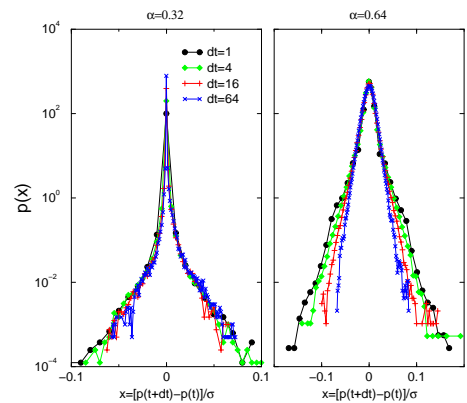,width=6cm,angle=0}}
\caption{Probability density function (pdf) of returns (rescaled to
unit variance) for different time lags $dt=1$, $4$, $16$ and $64$.
The market is composed of $N_s=100$ speculators with dynamic capitals 
($\epsilon=0.1$) and $N_p=10$ producers ($v_i=1$). Speculators
can decide not to play and those who loose all their capital are
replaced by new agent with random strategies and $c_i=1$.
$\alpha=0.32$ (left) and $0.64$ (right).}
\label{figpdfev}
\end{figure}

The dynamics of $c_i(t)$ gives an evolutionary character to the model,
because poorly performing agents are driven out of the market. Indeed,
asymptotically, a finite fraction of agents end up with
$c_i=0$. Evolutionary selection in the market can be introduced
assuming that agents with $c_i<\bar w\ll 1$ are replaced by new
agents, which enter with an initial capital $c_i=1$ and random trading
strategies. A further modification of the model lies in removing the
unrealistic feature of forcing agents to trade at each time step.  It
seems reasonable to allow agents not to trade, if their trading
strategies perform poorly\footnote{This is accomplished by assigning
to each speculator a special strategy -- called the $0$-strategy --
which prescribes not to trade ($a_i=0$), whatever the information
$\mu$ is.} \cite{JohnLiege}.  The phase transition separating an information efficient
phase $\alpha<\alpha_c$, from an inefficient phase $\alpha>\alpha_c$
survives to all these modifications\footnote{The value of $\alpha_c$
is non-universal, i.e. it depends on the parameters of the model.}.
Fig. \ref{figpdfev} shows that the rescaled pdf of returns on
different time lags $\Delta t$ collapse quite well close to
$\alpha_c\approx 0.32$ whereas a clear crossover to Gaussian behavior
occurs for $\alpha=0.64$. In other words, the crossover to a Gaussian
distribution of the distribution of returns $p(t+\Delta t)-p(t)$
occurs for a characteristic time lag $\Delta t_0$ which increases as
one approaches the critical point $\alpha_c$. This is reminiscent of
critical phenomena in statistical physics \cite{Stanley} where
correlation length and times diverge as the distance $\alpha-\alpha_c$
to the critical point vanishes.  In this framework of critical
phenomena, the crossover to a Gaussian pdf manifests itself as a
finite size scaling phenomenon.  Hence a measure of crossover times
$\Delta t_0$ in real markets allows one to estimate the parameter
$\alpha$, or its distance to criticality, in that market. This calls
for a systematic study of the relation between $\Delta t_0$ and
$|\alpha -\alpha_c|$ which goes beyond the scope of the present paper
and shall be discussed elsewhere.

It is also tempting to speculate that this relation between $\alpha$
and time-scales tells us how the number of market-relevant events over
a time window $\Delta t$ increases as the window size $\Delta t$
increases. At $\alpha_c$, all the original $P$ events have lost their
information content, hence the market is invariant under time
rescaling. At $\alpha=\alpha_c+\epsilon$, the unexploited information
remaining in the market is amplified by time rescaling. In other words,
that information becomes more and more detectable on larger and larger
time-scales. This is consistent with Figs. 3, 4 and 5, which show
that the market at longer 
and longer time-scales looks less and less critical\footnote{The 
opposite limit of high frequencies suggests
even more tempting speculations: The singularities arising in this
limit are reminiscent of those arising from quantum field theories of
interacting particles. This similarity suggests that renormalization
group approaches, a technique for studying scale-free systems,
 may be helpful to explain interacting markets at high
frequencies.}.

The crossover times to Gaussian behavior can be measured in real
markets. Its relation to the distance $|\alpha -\alpha_c|$ from the
critical point may serve as a basis to classify real markets according
to their distance from criticality.

\section{Market's ecology}

The artificial financial market described by the Minority Game can be
regarded as an ecosystem where different types of species of traders
interact. The three main species are {\em producers} -- who trade in a
deterministic way -- {\em speculators} -- who are adaptive -- and {\em
noise traders} -- who behave randomly (see the appendix).  The
interaction between these three species, which has been first studied
in ref. \cite{MMM}, will be the subject of the present section.

We shall discuss the Minority Game with fixed capitals, for which we
can rely on analytic results \cite{CMZe}. This allows us to quantify
the effect of the change in concentration of one species on itself,
on the other species and on the global behavior. 

As a measure of the efficiency of the market, one can take the {\em
signal to noise ratio}, which in the present context is just
$H/(\sigma^2-H)$. This accounts for the fact that even if some
profitable trading opportunities exist (i.e. $H>0$), they can only be
detected if their intensity exceeds that of stochastic fluctuations
(volatility $\sigma^2-H$). The signal to noise trader gives a measure
of efficiency which is particularly relevant for speculators.  A
second measure of efficiency is volatility: Market participants take
into account expected payoffs and risk, in a proportion related to
their risk aversion (and their time horizon). While speculators or
noise traders may be close to risk neutral, producers are risk
averse; for the latter, the fluctuations are a more relevant measure
of efficiency.

\begin{figure}
\centerline{\psfig{file=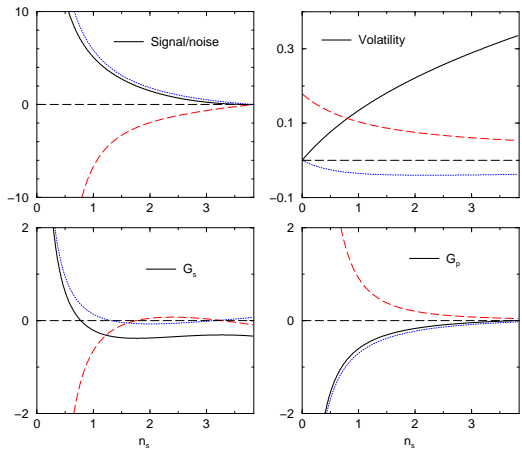,width=6cm,angle=270}}
\caption{Signal to noise ratio (upper left), volatility (upper right)
and payoffs of agents (lower panels; left for speculators and right
for producers) for a market with $n_p=1$ as a function of the (reduced)
number $n_s$ of speculators. Each graph also shows the variation of the
corresponding quantities if the number of producers (dotted lines) or
speculators (dashed lines) increase. The figures refer to the asymmetric
phase $H>0$ ($n_s<n_s^*$).}
\label{figecol}
\end{figure}

The impact of noise traders on the market ecology, as discussed
elsewhere\cite{MMM}, is easy to characterize. They do not contribute
to $H$, but they contribute to the losses $\sigma^2$.  Noise traders
do not affect the payoffs of other species. They only contribute to
volatility (see the appendix). We shall than concentrate on the
interaction between speculators and producers.  Fig. \ref{figecol}
illustrates the effects of adding an agent to a market with a fixed
number of producers, as the number of speculators varies (both numbers
are computed in units of $P$, see appendix).

\begin{figure}
\centerline{\psfig{file=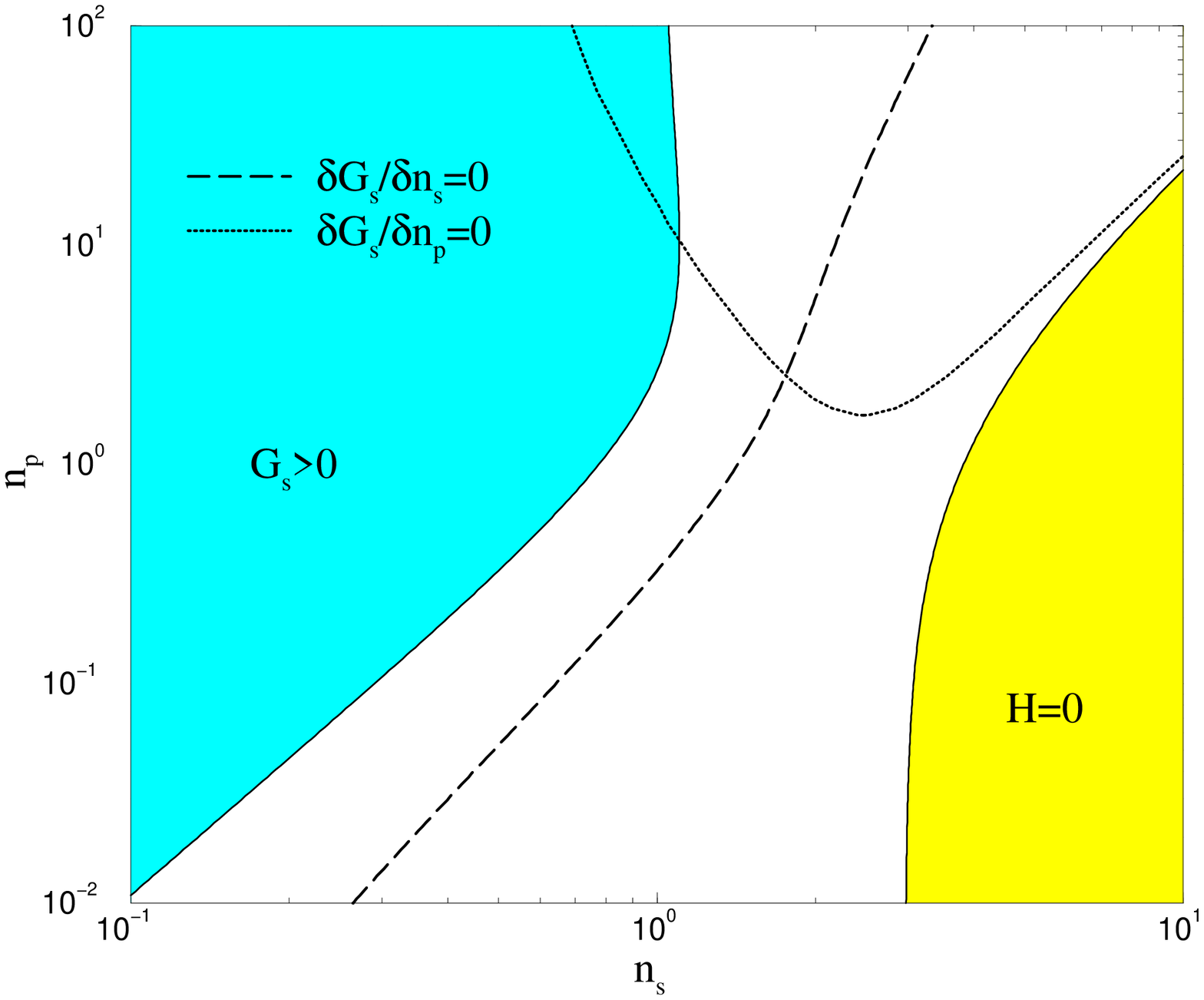,width=6cm,angle=270}}
\caption{Phase diagram of the MG in the $(n_s,n_p)$ plane. The symmetric
phase, where $H=0$, is the shaded area in the lower right corner. 
The payoff $G_s$ to speculators is positive in the upper left shaded
region. Above (below) the dotted line, an increase in the number $n_p$
of producers causes a decrease (increase) in the payoff $G_s$ of
speculators: i.e. $\partial G_s/\partial n_p <0$ (resp. 
$\partial G_s/\partial n_p >0$). The dashed line separates a 
region (left) where speculators are in competition (i.e. 
$\partial G_s/\partial n_s <0$), from a region (right) where $G_s$
increases with $n_s$.}
\label{nsnp}
\end{figure}

The signal to noise ratio decreases if new speculators enter the
market and increases as the number of producers increases. The
volatility instead increases with the number of speculators and
decreases with the number of producers: As the number of speculators
increases, the market becomes less predictable and speculators
themselves are less and less efficient in exploiting the information
present in the market. This results in the increase of volatility.  On
the other hand, increasing the number of producers, makes speculators
behave more efficiently.  The payoff of producers, which is always
negative, increases with the number of speculators and it decreases
with the number of producers themselves. This suggests that the
relationship between these two species may be better described as
symbiosis than as competition. Indeed generally also the payoff of
speculators increase if the number of producers increases. But as
Fig. \ref{figecol} shows, the situation for speculators is more
complex than that: If $n_p$ is large enough (i.e. above the dotted line
in fig. \ref{nsnp}), the gain of speculators
decreases if a other producers enter the game. Furthermore, close to
the boundary $n_s^*(n_p)$ of the symmetric phase, the gain of
speculators increases if a new speculator is added. This suggests that
the relationship among speculators cannot be described as competition
in this region (below the dashed line in fig. \ref{nsnp}). 
The phase diagram in the plane $(n_s,n_p)$, shown in fig. \ref{nsnp},
summarizes this behavior. 

These surprising 
result highlights the complexity of the interacting market system
described by the minority game.  The advantage of the Minority Game
with respect to other agent based models, is that this complexity can
be investigated analytically in detail, for simple cases.

\section{Conclusions}

The MG is not just a toy model, but a rather good starting point for 
modeling markets. By removing one by one its unrealistic features,
one obtains little by little stylized facts like fat tails and
algebraic decay of the volatility auto-correlation function; in
addition, the correspondence between stylized facts and
additional features put into the MG is very instructive.

But the MG is not only able to reproduce stylized facts. It is an
extremely powerful tool to explore the interplay between different
types of agents, and efficiency --~which can be well defined in this
model. The measure of efficiency should depend on which
type of agents is considered~: for instance, speculators are likely to
be interested in the signal to noise ratio, whereas producers are more
concerned with fluctuations. 

Information, the price mechanism and agents behavior in real markets
may be very different from those assumed in the MG.  However, if the
collective behavior of the market is due to statistical laws, we
expect it to be largely independent of microscopic details. From this
point of view, we expect the MG can say something about real markets.
For example, the phase transition from symmetric (unpredictable) to
asymmetric (predictable) markets is a very robust feature of MG's.
On one hand we expect a similar transition also in real markets,
on the other we showed how the distance from the critical point can be 
estimated. 

Further efforts to {\em calibrate} the MG to reproduce the statistical
features of a given market are certainly necessary to pursue this line
of research.

\appendix

\section{Definition of the MG}

Let ${\cal N}$ be the set of agents engaged in the minority game and
$N=|{\cal N}|$ be their number.
At each time $t$ each agent $i\in{\cal N}$ takes an action $a_i(t)$,
which is a real number quantifying his individual demand. 
The market interaction is defined in terms of the
``excess demand'' at time $t$, which is:
\be
A(t)=\sum_{i\in{\cal N}} a_i(t)
\ee
The ``volume'' of trades is defined as
\be
V(t)=\sum_{i\in{\cal N}} |a_i(t)|.
\ee
Let the return, at time $t$, be $r(t)=-A(t)/V(t)$ so that the payoffs to
each agent $i\in{\cal N}$ is
\be
g_i(t)=a_i(t)r(t)=-\frac{a_i(t)A(t)}{V(t)}.
\ee
This structure of interaction has the minority nature discussed in the
text: If $A(t)>0$ it is convenient to choose $a_i(t)<0$ and {\em vice-versa}.

As in refs. \cite{Caldarellietal,FarmerEco,bouchaudcont,JohnLiege,FarmerJoshi}, 
we define a price process by: 
\be\label{pricedyn}
\log p(t+1)=\log p(t)+r(t)=\log p(t)-\frac{A(t)}{V(t)}.
\ee
These equations are also the simplest ones dictated by dimensional analysis: 
$g_i$ has the same units as excess demand $A$ and return $r$ and $\log p$ 
are dimensionless (note that in ref. \cite{MMM} $g_i$ was not normalized
to the number of agents).

Agents observe a public information, which can take one of $P$ forms, 
labeled by an integer. $\mu(t)$ is the {\em information} at time 
$t$, which we assume here to be randomly drawn at each time\footnote{Much
of what follows can be extended to the original case where $\mu(t)$ 
encodes the sign of $A(t')$ in the last $\log_2 P$ periods of the game
 \cite{CM00}.}. We distinguish three types of agents according to 
their behavior with respect to information: 
\be\label{typtr}
a_i(t)=\left\{
\begin{array}{cc}
{\rm rand} & i\in{\cal N}_n\\
v_i\sigma^{\mu(t)}_i & i\in{\cal N}_p\\
w_i\sigma^{\mu(t)}_{i,s_i(t)} & i\in{\cal N}_s
\end{array}\right.
\ee
The first type of agents (${\cal N}_n$) shall be called {\em noise
traders}. They totally disregard information and take actions at
random (i.e. with no correlation to $\mu(t)$). For example $a_i(t)=+1$
or $-1$ with equal probability. The second type, called {\em
producers} (${\cal N}_p$), behaves in a deterministic way, given
$\mu(t)$.  $v_i$ is the amount they invest in the market and
$\sigma_i^\mu$ is a random function of $\mu$ into $\{\pm 1\}$, drawn
independently for each $i\in {\cal N}_p$. These functions are called
{\em strategies} for short, but producers do not optimize their 
behavior: they only have one strategy. {\em Speculators}, which
are the third type of traders (${\cal N}_s$ in Eq. (\ref{typtr}), 
instead can optimize their behavior dynamically: They have $S$ strategies,
$\sigma_{i,s}^\mu$ labeled by the index $s=1,\ldots,S$, and 
can choose the one which performs better, by adjusting $s$\footnote{This
is done by assuming that agents assign scores $U_{i,s}(t)$ to each of their
strategies. Scores are updated according to the virtual performance of a
strategy $s$: $U_{i,s}(t+1)=U_{i,s}(t)+w_i\sigma_{i,s}^{\mu(t)}r(t)$.
See \cite{MCZ,MC} for a discussion of issues related to this type
of learning.}. Strategies
are again drawn randomly and independently for each $i$ and $s$.
The amount $w_i$ invested by speculator $i$ will be discussed below.
Hence ${\cal N}={\cal N}_n\cup{\cal N}_p\cup{\cal N}_s$ and
$N=N_n+N_p+N_s$ where $N_x={\cal N}_x$ is the size of population 
of type $x=n,p$ or $s$. The MG has been introduced with $N_n=N_p=0$
and $w_i=1$, $\forall i\in{\cal N}_s$. The case $N_n>0$, $N_p>0$ has
been first discussed in ref. \cite{MMM}, always with $w_i=v_i=1$.
For these cases an analytic solution in the limit $N\to\infty$ has
been found with $P/N_s=\alpha$, $N_p/N_s=\rho$ and $N_n/N_s=\eta$ 
fixed. Rather than using these parameters, we prefer to discuss
our results in the rescaled population variables:
\be
n_n=\frac{N_n}{P}=\frac{\eta}{\alpha},
~~~n_p=\frac{N_p}{P}=\frac{\rho}{\alpha},~~~n_s=\frac{N_s}{P}
=\frac{1}{\alpha}.
\ee

The key quantities of interest are 
\be
\sigma^2=\ovl{\avg{A^2}}=\frac{1}{P}\sum_{\mu=1}^P \avg{A^2|\mu},
\ee
which is proportional to the total losses of agents $\sum_{i\in {\cal N}} 
\avg{g_i}=-\sigma^2/V$, and 
\be
H=\ovl{\avg{A}^2}=\frac{1}{P}\sum_{\mu=1}^P \avg{A|\mu}^2
\ee
which measures the predictability of the market's outcome $A(t)$.
Here and below the average over $\mu$ is denoted by an overbar and
the time average, conditional to $\mu(t)=\mu$ is denoted by
$\avg{\cdot|\mu}$. 

\subsection{Analytic solution with $w_i=v_i=1$}

The MG with speculators, producers and noise traders and fixed
capitals, has been studied in ref. \cite{MMM}. We refer the interested
reader to that work and report here only the final expressions of 
the analytic solution \cite{MMM} in terms of
the parameters $n_x$: 
\be 
H=P[1-n_s\erf(z)]^2
\left[\frac{1+Q}{2}n_s+n_p\right] 
\ee 
where $n_x=N_x/P$ for $x=s,p,n$
are reduced concentrations. $Q$ is a function of $z$ 
\be
Q=1-\frac{e^{-z^2}}{\sqrt{\pi}z}-\left(1-\frac{1}{2z^2}\right)\erf(z)
\ee 
and $z=z(n_s,n_p)$ is the solution of the equation 
\be
2[(1+Q)n_s+2n_p]z^2=1.  
\ee 
Note that $H$ only depends on $n_s$ and
$n_p$. Noise traders have no effect on it. Ref. \cite{MMM} finds
\be
\sigma^2=H+P \left(\frac{1-Q}{2} n_s +n_n \right).
\ee
The payoff of producers is
\be
G_p=\sum_{j\in {\cal N}_p} \avg{g_j}
=-\frac{n_p}{n_s+n_p+n_n}\left[\frac{1}{n_s}-\erf(z)\right]
\ee
and that of noise traders is simply $G_n=-n_n/(n_s+n_p+n_n)$.
The payoff of speculators is then $G_s=-\sigma^2/N-G_p-G_n$ with
$N=N_s+N_p+N_n$. These expressions where used to produce the results 
in the text.

\subsection{MG with heterogeneous weights of agents}

The analytic solution generalizes easily to the case where 
the weights $w_i$ of the agents are randomly drawn from a given
pdf $P(w)$ at the beginning of the game and kept fixed afterwards.
For simplicity we deal with the case $n_p=n_n=0$ and 
$n_s=1\alpha>0$, though other cases 
are easily dealt with.
Without loss of generality, we can fix the ``scale'' of $A(t)$ by
imposing that the average wealth is $\avg{w_i}=1$ (a different value
of $\avg{w_i}$ is restored by dimensional analysis). If $\avg{w_i^2}<\infty$,
following the same calculation as in ref.\cite{CMZe,MCZ,MMM}, we find that
\be 
\label{H_c}
H=P\avg{w^2}_w(1+Q)\left[1-n_s\avg{\erf(wz)}_w\right]^2
\ee
where $\avg{\cdots}_w$ stands for averages over the distribution $P(w)$,
\be
Q=1-\frac{\avg{w e^{-w^2 z^2}}_w}{\sqrt{\pi}\avg{w^2}_wz}-
\left\langle\left(w^2-\frac{1}{2 z^2}\right)\frac{\erf(wz)}{\avg{w^2}_w}
\right\rangle_w
\ee
and $z$ is the solution of the equation
\be
2z^2\avg{w^2}_w(Q+1)=1.
\ee
The order parameter $Q$ is defined, in this case, as
\be
Q=\frac{1}{N\avg{w^2}_w}\sum_{i=1}^N w_i^2 m_i^2
\ee
where $m_i$ is the ``local magnetization'' of agent $i$, i.e. the
excess probability with which $i$ plays the strategy $s_i=+1$.
Also $\sigma^2=H+P\avg{w^2}_w(1-Q)/2$. The phase transition occurs for 
a critical $n_s^\star$ such that $n_s^\star\avg{\erf(wz)}_w=1$.

These equations hold as long as $\avg{w^2}_w$ is finite. When the
second moment of $w_i$ diverges, i.e. when $P(w)\sim w^{-\gamma-1}$
for $w\gg 1$ with $\gamma<2$, one expects large fluctuations and no
self-averaging.  Indeed sums of the form $\sum_i w_i^2 (\cdots)$,
which define order parameters, are dominated by the richest agent and
scale with $N$ faster than linearly ($\sum_i w_i^2\sim N^{2/\gamma}$).
These sums do not satisfy laws of large numbers and the
rescaled variable $N^{-2/\gamma}\sum_i w_i^2(\ldots)$ does not converge
to a constant, as in the law of large numbers, but rather fluctuates 
for all $N$. Standard statistical mechanics approaches
breaks down in these cases.

\section{appendix: MG with dynamical capitals}

In real markets, the weight of each agent is not a fixed
quantity, for instance because her capital evolves in
time.  Indeed, poorly performing speculators will eventually be
ruined and will not participate to the market. We generalize the
MG in order to account for this very important fact. Each speculator
$i\in{\cal N}_s$ has a capital
$c_i(t)$ and invests a fraction $\epsilon$ of it in the market:
Hence $w_i(t)=\epsilon c_i(t)$. The capital of a speculator evolves
in time according to his performance:
\be
c_i(t+1)=c_i(t)+g_i(t)=c_i(t)[1+\epsilon \sigma_{i,s_i(t)}^{\mu(t)}r(t)]
\ee
If agent $i$ loses ($g_i(t)<0$) his capital decreases and {\em vice-versa}.

Without producers, the gain of the speculators is always negative and
hence the total capital of speculators decreases and tends to
zero. When producers are present, the total capital of speculators
adjusts so that speculators have $\avg{g_i}=0$ and a stationary state
is possible. This is in principle accessible to an analytic
calculation \cite{CMZe} for $\epsilon\ll 1$. In this case indeed one
can rely on an {\em adiabatic} approximation where strategies adjust
instantaneously to any small change in capitals $c_i(t)$.
This implies that one may consider $c_i(t)$ as ``quenched disorder'' (as in
the previous calculation) and impose, self-consistently that 
$\log c_i(t)$ is a stationary process (i.e. $\avg{\log [c_i(t+1)/c_i(t)]}=0$).
Though feasible, this approach involves quite complex calculations.

\end{document}